\documentclass[preprint,10pt]{sigplanconf}

\usepackage{times}
\usepackage{amscd}
\usepackage{amsmath}
\usepackage{url}
\usepackage{graphicx}
\usepackage{lscape}
\usepackage{listings}
\lstMakeShortInline[language=C]!

\usepackage{times}
\SetMathAlphabet{\mathtt}{normal}{OT1}{pcr}{m}{n}
\SetMathAlphabet{\mathtt}{bold}{OT1}{pcr}{bx}{n}
\usepackage{amsmath}
\usepackage{amssymb} % for \pitchfork

\newcommand{\CUT}[1]{}

\newcommand{\appref}[1]{Appendix~\ref{#1}}

\newcommand{\secref}[1]{Section~\ref{#1}}
\newcommand{\tblref}[1]{Table~\ref{#1}}
\newcommand{\figref}[1]{Figure~\ref{#1}}

\newcommand{\eg}{{\em e.g.}}

\newcommand{\ie}{{\em i.e.}}

\newcommand{\nesl}{{\textsc{Nesl}}}

\newcommand{\cml}{\textsc{CML}}
\newcommand{\pml}{\textsc{PML}}

\providecommand{\bftt}[1]{{\ttfamily\bfseries{}#1}}

\providecommand{\kw}[1]{\bftt{#1}}
\providecommand{\nt}[1]{{\rmfamily\itshape{#1}}}

\newcount\timeHH
\newcount\timeMM
\timeHH=\time
\divide\timeHH by 60
\timeMM=\time
\multiply\count255 by -60 \advance\timeMM by \count255
\newcommand{\timestamp}{%
  \today{} ---
  \ifnum\timeHH<10 0\fi\number\timeHH\,:\,\ifnum\timeMM<10 0\fi\number\timeMM}

\newcommand{\vproc}{vproc}

\usepackage{common/code}

\title{Garbage Collection for Multicore NUMA Machines}
\authorinfo{Sven Auhagen}{%
  University of Chicago}{%
  sauhagen@cs.uchicago.edu}
\authorinfo{Lars Bergstrom}{%
  University of Chicago}{%
  larsberg@cs.uchicago.edu}
\authorinfo{Matthew Fluet}{%
  Rochester Institute of Technolgy}{%
  mtf@cs.rit.edu}
\authorinfo{John Reppy}{%
  University of Chicago}{%
  jhr@cs.uchicago.edu}

\conferenceinfo{MSPC'11,} {June 5, 2011, San Jose, California, USA.}
\CopyrightYear{2011}
\copyrightdata{978-1-4503-0794-9/11/06}

\begin{document}

\maketitle
\thispagestyle{empty}

\sloppy

%!TEX root = paper.tex
%
\begin{abstract}
Modern high-end machines feature multiple processor packages, each of which
contains multiple independent cores and integrated memory controllers connected
directly to dedicated physical RAM.
These packages are connected via a shared bus, creating a system with a
heterogeneous memory hierarchy.
Since this shared bus has less bandwidth than the sum of the links to memory,
aggregate memory bandwidth is higher when parallel threads all access memory
local to their processor package than when they access memory attached to a
remote package.
This bandwidth limitation has traditionally limited the scalability of modern
functional language implementations, which seldom scale well past 8~cores, even
on small benchmarks.

This work presents a garbage collector integrated with our strict, parallel
functional language implementation, Manticore, and shows that it scales
effectively on both a 48-core AMD Opteron machine and a 32-core Intel Xeon
machine.
\end{abstract}




%!TEX root = paper.tex

% SMP to NUMA picture -- limits should be no surprise
% GC overview

\section{Introduction}
Inexpensive multicore processors and accessible multiprocessor motherboards
have brought all of the challenges inherent in parallel programming with large numbers of
threads with non-uniform memory access (NUMA) into the foreground.
Functional programming languages are a particularly interesting approach
to programming parallel systems, since they provide a high-level programming
model that avoids many of the pitfalls of imperative parallel programming.
But while functional languages may seem like a better fit for parallelism due to
their ability to compute independently while avoiding race conditions and
locality issues with shared memory mutation, implementing a scalable functional
parallel programming language is still challenging.
Since functional languages are value-oriented, their performance is highly
dependent upon their memory system.

Our group has been working on the design and implementation of a
parallel functional language to address the opportunity afforded by
multicore processors.
In this paper, we focus on the design of our memory system and parallel garbage 
collector.
This system is designed to minimize required synchronization and to maximize
locality, two features which have proven crucial to the scalability of our
system on larger machines.
Recent work on other functional languages has shown that the memory system is
the limiting factor to improved performance for many types of
code~\cite{multicore-haskell,intel-private-heap}.
Our work has been guided by measurements of a number of parallel
benchmarks; we present detailed results from a representative
subset of these programs.

This paper makes the following contributions:
\begin{enumerate}
  \item
    We demonstrate a modern functional language that makes effective use of a
    large number of modern NUMA multicore processors.
    The best recent work scales to no more than 12~cores, and we demonstrate
    good utilization of all available cores on both 32~and~48~core machines.
    This scaling is demonstrated through a set of small but representative
    benchmarks across a variety of physical memory allocation strategies.
  \item
    We describe our garbage collector, which provides excellent performance
    on multicore, NUMA machines.
    While some of the individual ideas in the garbage collector build on classic
    work~\cite{appel:simple-gc, concurrent-caml-gc, portable-multiprocessor-gc},
    we present a novel approach that, when combined with other aspects of our runtime
    architecture designed to maximize locality, avoids bottlenecks due to
    excessive memory traffic. 
\end{enumerate}%

The remainder of the paper is organized as follows.
In the next section, we describe our language and runtime system.
\secref{sec:collector} lays out the architecture of our garbage collector.
\secref{sec:evaluation} contains a detailed evaluation of our implementation
using some representative benchmarks.
Due to length constraints, a discussion of related work is omitted.

%%% Local Variables: 
%%% mode: latex
%%% TeX-master: "paper"
%%% End: 

%!TEX root = paper.tex

\section{Manticore overview}
\label{sec:manticore}

The Manticore project encompasses both design and implementation of
parallel functional programming languages on modern multicore and
multiprocessor systems.
In this section, we give a brief overview of the features relevant to
threading and the garbage collector.
More detail can be found in our previous
papers~\cite{manticore-damp07,manticore-ml07}.

\subsection{Programming model}
Parallel ML (PML) is the programming language supported by the Manticore
system.
Our programming model is based on a strict, but mutation-free, functional
language (a subset of Standard ML~\cite{sml97-definition}), which is extended
with support for multiple forms of parallelism.
This subset includes most of the core features of SML as well as a simple module
system. 
PML differs from SML primarily by lacking mutable data (\ie{}, reference cells
and arrays), but it does include exceptions. 
\pml{} extends this sequential core with both fine-grained implicitly-threaded
and coarse-grain explicitly-threaded~\cite{parallel-cml} parallel-programming mechanisms.
The implicitly-threaded mechanisms include a variety of lightweight
syntactic forms that allow the programmer to suggest to the compiler
and runtime system that parallelism would be
beneficial~\cite{implicit-threading-in-manticore}; because the threads
used to evaluate these constructs are not visible at the language level, the constructs are
termed \emph{implicitly threaded}.
The explicitly-threaded mechanisms include language-level visible
threads and synchronous message passing, providing a parallel
implementation of Concurrent~ML's concurrency
primtives~\cite{parallel-cml}.

\subsection{The Manticore runtime system}
The Manticore runtime system consists of a hardware abstraction level, which is
written in C, that supports \emph{virtual processors} (\vproc{}), basic system
services, such as I/O and networking, and a parallel garbage collector.
A \vproc{} is an abstraction of a computational resource, and is used
to execute code and balance work across the system. 
Each \vproc{} is hosted by its own pthread~\cite{butenhof:pthreads-book}, which
is pinned to a physical node.
When there are less \vproc{}s than processors, they are assigned sparsely across
the nodes to minimize contention on the node-shared L3 cache.

\subsection{Execution of parallel work and locality}
\label{manticore:locality}
All of the implicitly threaded parallelism language features work by pushing units of
parallel work (in the form of continuations) onto a \vproc{}-local work queue
and then beginning execution of the first unit of work.
If a \vproc{} has no work to perform, then it uses \emph{work-stealing} to find
a unit of pending work on another \vproc{} and begins executing it.
This strategy is designed to keep memory and computation local to the thread
that began the work whenever possible and leads to one of the key invariants
provided by our runtime system and used by our garbage collector --- all data is
local to a processor unless it was either captured in a closure and stolen by
another processor or it is passed in a message by the \cml{} explicit threading
features.
At these two points, the runtime and basis library handle copying data out of
the local heap and into the global space, as we describe in
\secref{sec:collector:heap}.
This invariant means that:
\begin{enumerate}
\item There are no pointers from one \vproc{}'s local heap to another's.
\item There are no pointers from the global heap into any \vproc{}'s local
  heap.
\end{enumerate}%
Many related collectors require these properties to obtain concurrency or
parallelism.
Our approach differs from theirs by requiring neither write barriers nor static
analysis to maintain these properties.

%%% Local Variables: 
%%% mode: latex
%%% TeX-master: "paper"
%%% End: 

%!TEX root = paper.tex

\section{GC and heap}
\label{sec:collector}

Our garbage collector is based on a novel combination of the 
Doligez-Leroy-Gonthier (DLG) parallel
collector~\cite{concurrent-caml-gc,portable-multiprocessor-gc}
and the Appel semi-generational collector~\cite{appel:simple-gc}.
This design allows us to minimize GC synchronization between
vprocs and to preserve locality.

%The heap is organized into a fixed sized local heap for each \vproc{} and a
%shared global heap, which a collection of memory chunks.
%Each vproc has a dedicated memory chunk in the global heap.

\subsection{Heap architecture}
\label{sec:collector:heap}
We use the DLG heap architecture of per-vproc local heaps combined
with a global heap.
As in the DLG collector, we maintain the invariant that there are no pointers
between local heaps or from the global heap into the local heap.
This invariant means that for one vproc to communicate an object to another,
we must first \textit{promote} the object to the global heap.
The cost of promotion can be a significant burden, so we have developed
a number of techniques for reducing the amount of promoted data.
These include a lazy promotion scheme for work stealing~\cite{rainey:phd}
and the use of object proxies.\footnote{
  Object proxies are a special kind of object that is used to allow references from
  the global heap back into the local heap.
  We use them in the implementation of our explicit concurrency constructs.
}

Functional-language implementations are notorious for their high rate of memory
allocation.
Fortunately, most of this data is ephemeral and so generational techniques are
quite effective.
To this end, we use Appel's semi-generational heap architecture for the local heaps.
The local heaps are fixed size that is chosen so that the local heaps will fit into
the L3 cache.

The global heap is organized into a collection of chunks.
Each vproc has a current chunk that it uses when it needs to allocate in
or promote an object to the global heap.
In a NUMA system, each node has its own bank of memory with the property that
access from a node to its own memory is faster than access to memory on other
nodes.
For this reason, our memory system tracks the node on which a chunk is allocated
and preserves node affinity when reusing chunks.

The main advantage of the DLG split-heap architecture is that it requires little or no synchronization
between vprocs for most garbage collection activity.
Our system has three different garbage-collection phases: minor, major, and global.
The former two correspond to Appel's minor and major collections and are used to reclaim
space in the local heap.
The global collection is a parallel stop-the-world collector.
We describe these in more detail below.

\subsection{Object representation and scanning}
\begin{figure}[t]
  \begin{center}
    \includegraphics[scale=0.4]{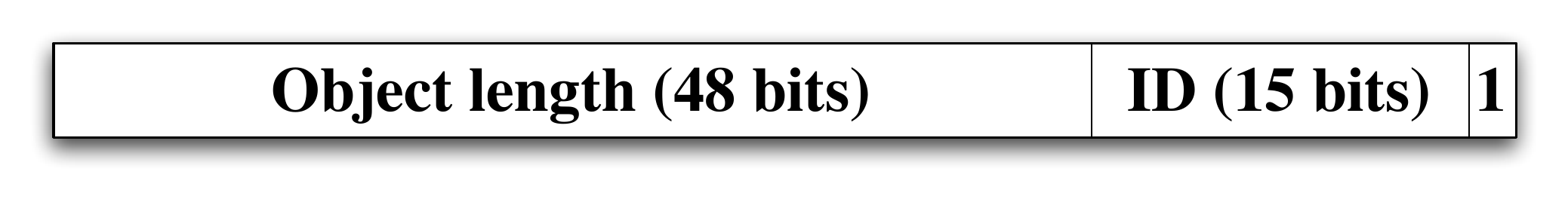}
  \end{center}%
  \caption{The header word of mixed-type, raw, and vector heap objects}
  \label{fig:headerword}
\end{figure}

The Manticore memory system supports three basic kinds of heap objects:
raw-data objects (\eg{}, strings), vectors of pointers, and mixed-type objects,
which contain both pointer and non-pointer data.
Heap objects have a 64-bit header word as shown in \figref{fig:headerword}.
The lowest bit is always 1, which distinguishes headers from forward pointers.
The rest of the header word consists of a 15-bit ID and a 48-bit length.
We reserve two IDs for raw and vector data.
For mixed objects, the ID is an index into an object-descriptor table that is
generated by the compiler.
The object-descriptor table includes pointers to object-scanning and
forwarding functions, which are also generated by the compiler.

Each garbage-collection function in the table is specifically created for the
structure of the corresponding mixed-type object.
This approach allows the garbage collector to avoid scanning each field of an object at runtime and instead to
generate code during compilation that processes only the pointer fields of each 
object.
We follow this approach for all mixed-type objects, though the garbage collector
still distinguishes raw and vector objects and handles them directly to avoid
a pointer lookup in the object table.

\subsection{Minor and major collections}

\begin{figure}[t]
  \begin{center}
    \includegraphics[scale=0.4]{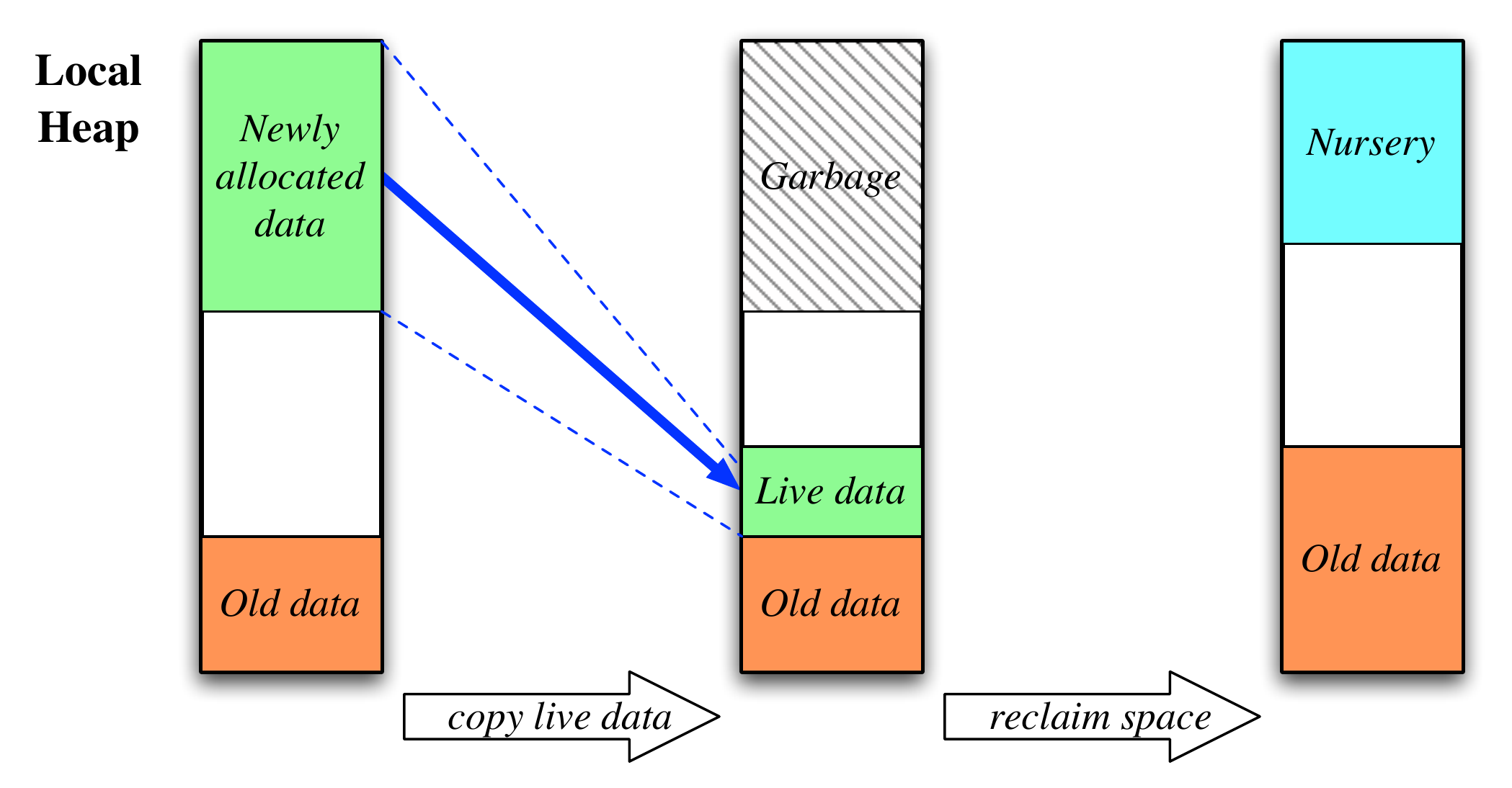}
  \end{center}%
  \caption{A minor garbage collection in Manticore}
  \label{fig:minor-gc}
\end{figure}

Following Appel, we divide a \vproc{}'s local heap into two separate spaces: the nursery area and the old-data area. 
New objects are allocated in the nursery area until it is full and a minor garbage
collection is triggered.
The minor garbage collector copies all live data from the nursery area to the old-data area of
the local heap.
After this minor garbage collection finishes, the remaining free space in the local heap is divided in half and the
upper half will be used as the new nursery area.
This process is illustrated in \figref{fig:minor-gc}.
Because there are no pointers into the local heap from outside (other than the roots),
minor collections require no synchronization.
A minor garbage collection triggers a major garbage collection when the size of
the new nursery area falls below a certain threshold or if a
global garbage collection is pending.

\begin{figure}[t]
  \begin{center}
    \includegraphics[scale=0.4]{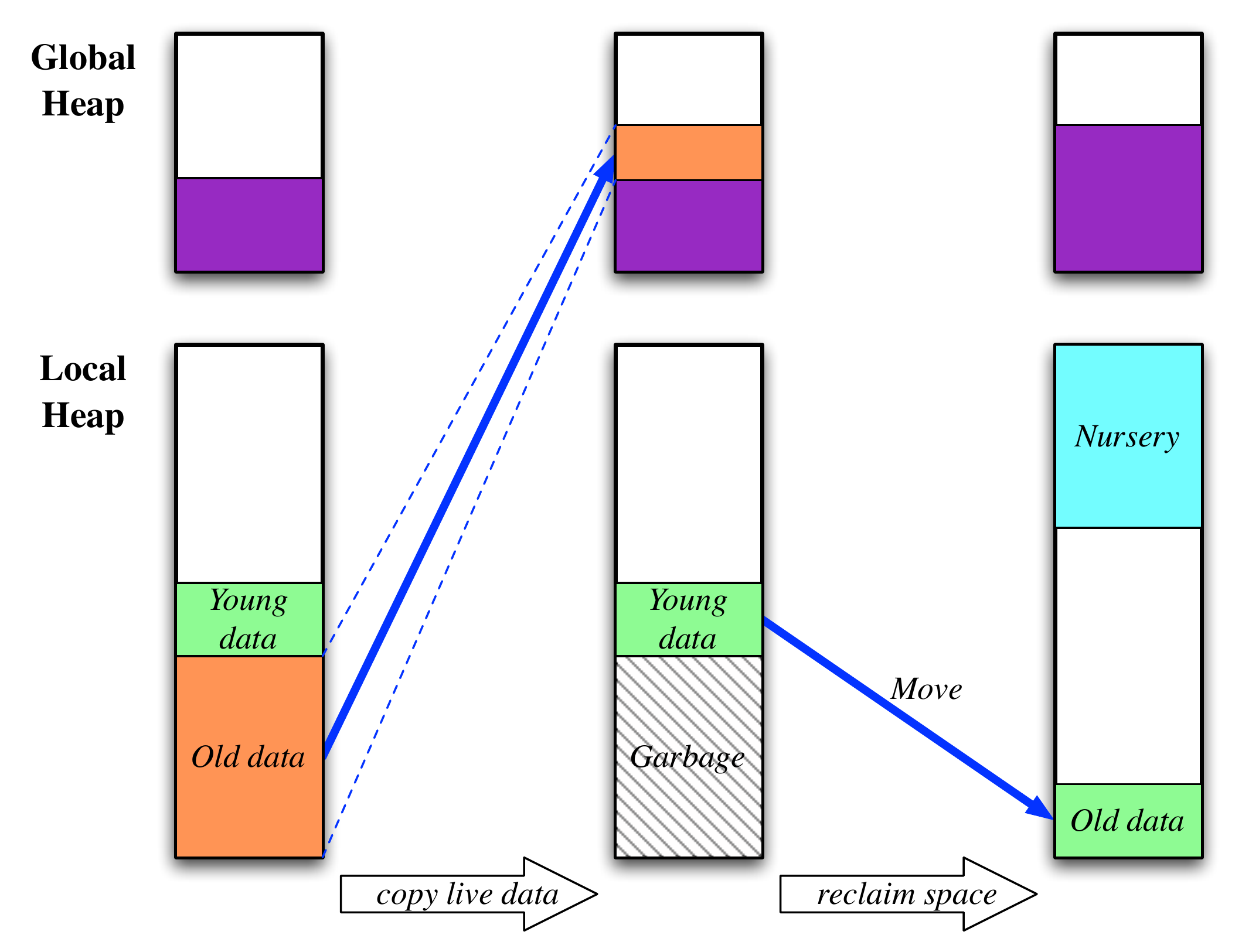}
  \end{center}%
  \caption{A major garbage collection in Manticore}
  \label{fig:major-gc}
\end{figure}

The major garbage collection copies the live objects from the old-data area in a
\vproc{}'s local heap to its dedicated memory chunk in the global heap.
To avoid premature promotion, we partition the old-data area into data that
was just copied in the previous minor collection (called young data) and the
data that was copied earlier.
The young data are guaranteed to be live (because a minor collection
always immediately precedes this major collection) and we do not copy
it to the global heap.
\figref{fig:major-gc} illustrates this process.

Major collections only require synchronization when the \vproc{}'s current memory
chunk is exhausted, since, in that case, the \vproc{} needs to allocate a
new chunk of global memory.
This synchronization is either node-local because it involves the reuse of a
chunk of memory or global if a new chunk needs to be requested from the system
and registered with the runtime.

In addition to minor and major collections, the runtime system also implements
object promotion, which is required when an object is to be shared with other
\vproc{}s.
Promotion is essentially a major collection, where the root set is a pointer
to the promoted object, and the synchronization requirements are the same as
for major collection.

\subsection{Global collection}

Global collection is triggered when the size of global heap chunks allocated
exceeds a threshold.\footnote{
  Currently, this threshold is the number of
  \vproc{}s times 32MB.
}
The \vproc{} that determines that a global collection first attempts to trigger
a global collection.
After the collection is triggered, one \vproc{} is assigned the leadership role
and performs the following actions.
\begin{enumerate}
\item Set a global flag that a global garbage collection is in progress and mark
  this \vproc{} the leader.
\item Signal all of the other \vproc{}s to enter garbage collection code by
  setting their allocation limit pointer to zero. This strategy allows the
  runtime to know that all \vproc{}s will be at a safe execution point with
  known roots.
\item Wait for all of the other \vproc{}s to enter the global collection, which
  requires first performing their parallel minor and major collections.
\end{enumerate}

At this point, every \vproc{} will be in the state shown at the end of
\figref{fig:major-gc}.
Everything pointed to by the roots and local heap will be present either
elsewhere in the local heap or in a global heap chunk.
These global heap chunks are gathered on a per-node basis and placed into a list
of from-space chunks.
Each \vproc{} then obtains a new global heap chunk and scans the \vproc{}'s
roots and local heap, placing any objects pointed-to into this new to-space
chunk.
In parallel with one another, the \vproc{}s obtain chunks on a per-node basis
from either the from-space list or the list of to-space chunks that have not
been scanned.
Each of these chunks are removed and scanned until no chunks remain on the local
node.
Once all of the \vproc{}s across all nodes have completed, the old
from-space chunks are returned to the free-space chunk pool and execution of the
program resumes.
\CUT{
A performance comparison between this parallel strategy and using only one
\vproc{} per node is in \secref{sec:evalGlobal}.
}

%%% Local Variables: 
%%% mode: latex
%%% TeX-master: "paper"
%%% End: 

%\input{numa}
%!TEX root = paper.tex

\section{Evaluation}
\label{sec:evaluation}

Our 32~core Intel and 48~core AMD hardware is described in detail in
\appref{sec:hardware}.

\subsection{Benchmarks}
For our empirical evaluation, we use five benchmark programs from our benchmark suite
and one synthetic benchmark.
Each benchmark is written in a pure, functional style and was originally written
by other researchers and ported to our system.
We ran each experiment 10 times and we report the average performance results in
our graphs and tables.

The Barnes-Hut benchmark~\cite{barnes-hut} is a classic N-body problem solver.
Each iteration has two phases.
In the first phase, a quadtree is constructed from a sequence of mass points.
The second phase then uses this tree to accelerate the computation of
the gravitational force on the bodies in the system.
Our benchmark runs 20 iterations over 400,000 particles generated in
a random Plummer distribution.
Our version is a translation of a Haskell
program~\cite{barnes-hut-haskell-bench}. 

The Raytracer benchmark renders a $512 \times 512$ image in parallel as
two-dimensional sequence, which is then written to a file.
The original program was written in ID~\cite{id90-manual} and is a simple
ray tracer that does not use any acceleration data structures.
The sequential version differs from the parallel code in that it
outputs each pixel to the image file as it is computed, instead of building
an intermediate data structure.

The Quicksort benchmark sorts a sequence of 10,000,000 integers in parallel.
This code is based on the \nesl{} version of the algorithm~\cite{scandal-algorithms}.

The SMVM benchmark is a sparse-matrix by dense-vector multiplication.
The matrix contains 1,091,362 elements and the vector 16,614.

The DMM benchmark is a dense-matrix by dense-matrix multiplication in which
each matrix is $600 \times 600$.

\begin{figure}
\begin{center}
  \includegraphics[width=3.2in]{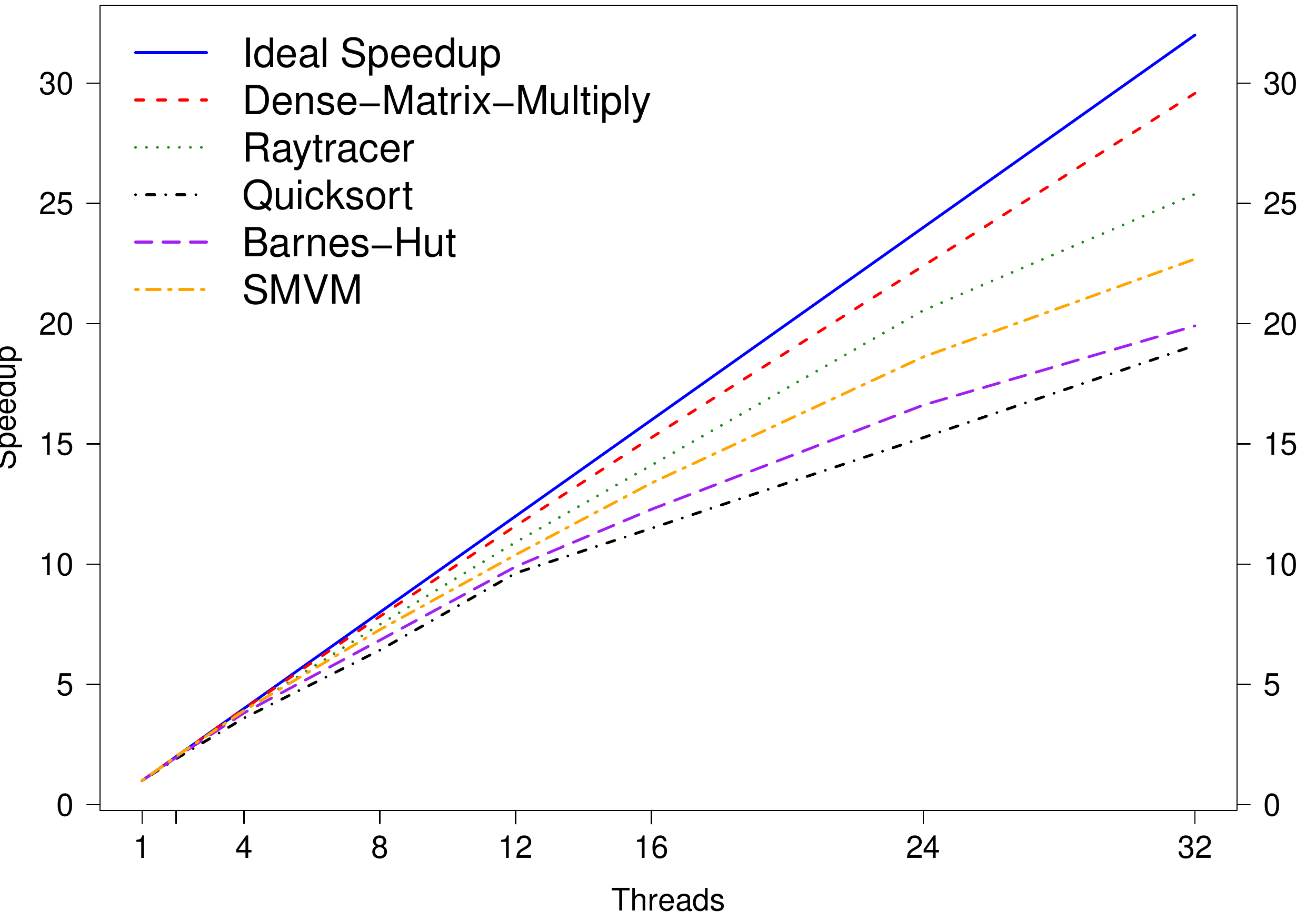}
\end{center}%
\caption{ 
  Comparative speedup plots for five benchmarks on Intel hardware.
}
\label{fig:intel-speedups} 
\end{figure}%

\begin{figure}
\begin{center}
  \includegraphics[width=3.2in]{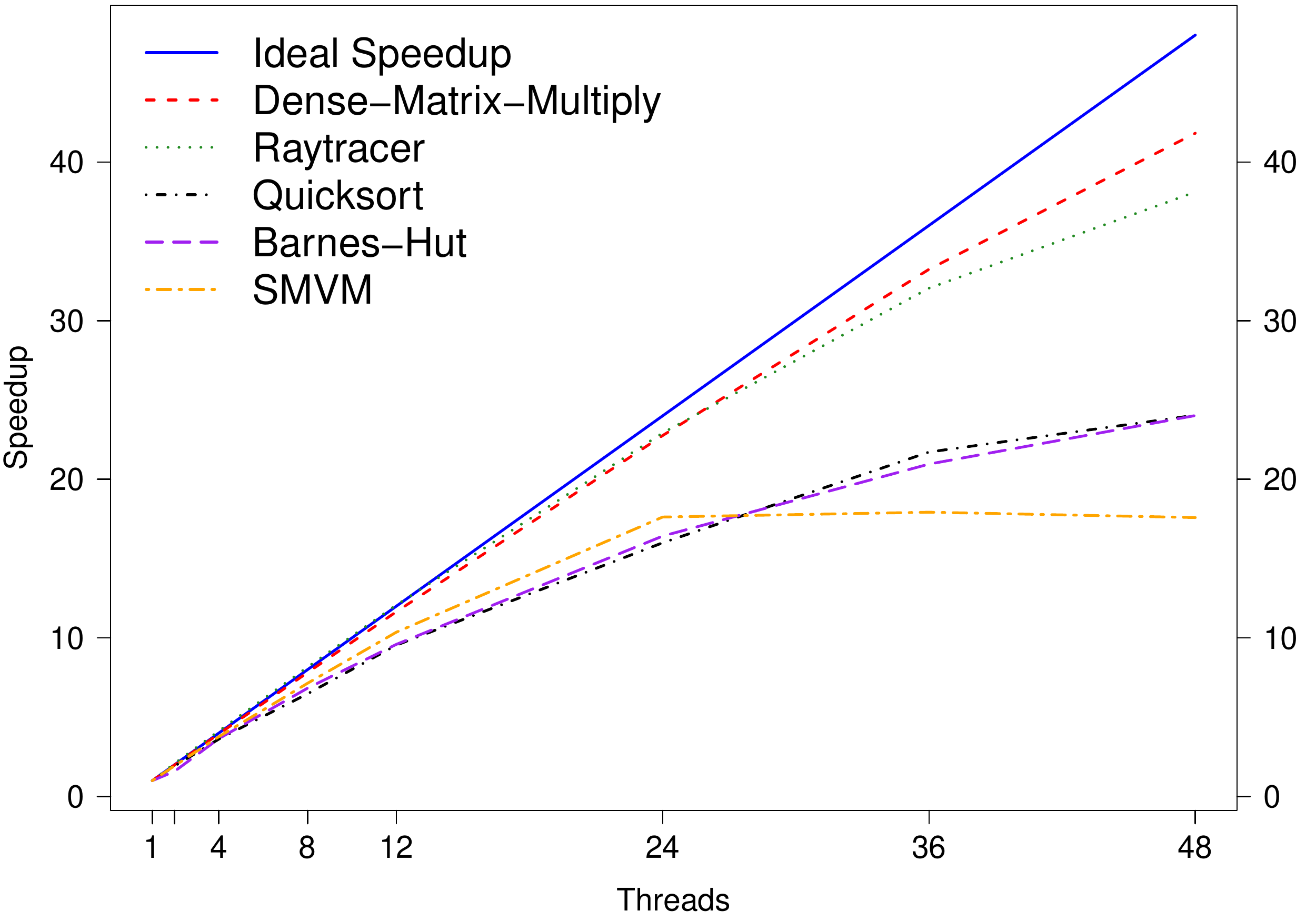}
\end{center}%
\caption{ 
  Comparative speedup plots for five benchmarks on AMD hardware using local
  memory allocation.
}
\label{fig:amd-speedups} 
\end{figure}%

\begin{figure}
\begin{center}
  \includegraphics[width=3.2in]{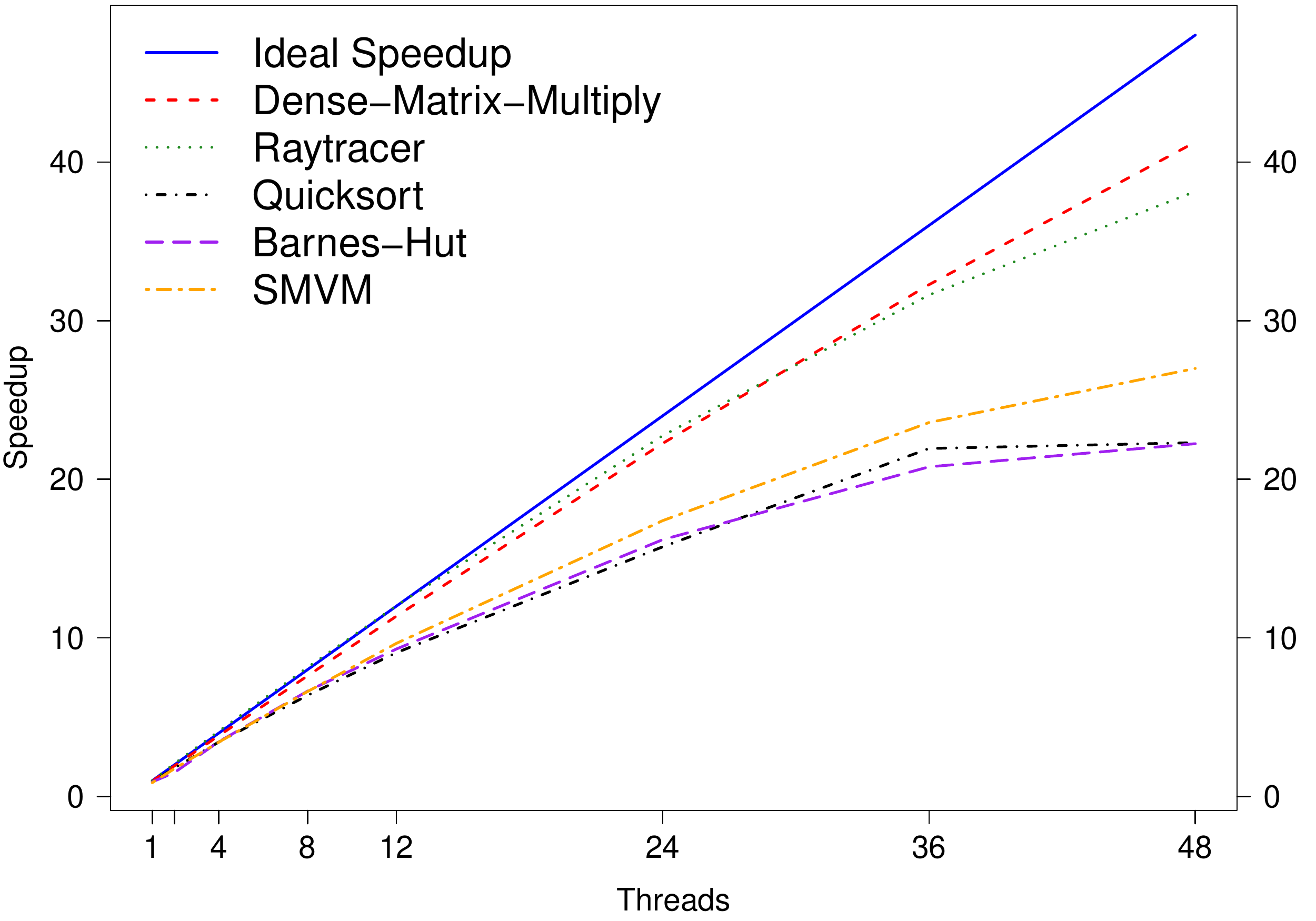}
\end{center}%
\caption{ 
  Comparative speedup plots for five benchmarks on AMD hardware with interleaved
  memory allocation.
}
\label{fig:amd-speedups-interleaved} 
\end{figure}%

\begin{figure}
\begin{center}
  \includegraphics[width=3.2in]{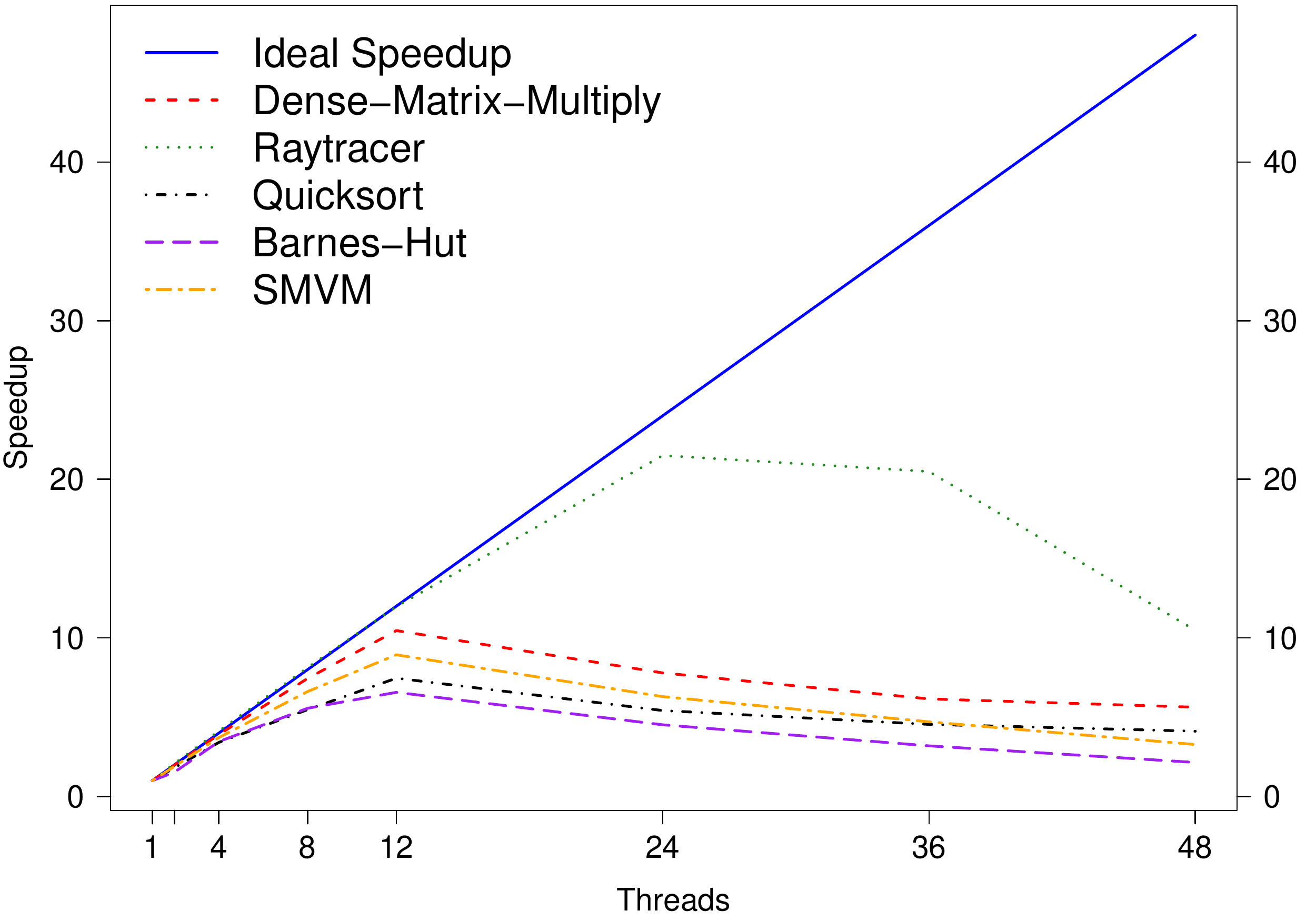}
\end{center}%
\caption{ 
  Comparative speedup plots for five benchmarks on AMD hardware with socket zero
  memory allocation.
}
\label{fig:amd-speedups-single} 
\end{figure}%

\subsection{Performance}

As shown in \figref{fig:intel-speedups}, on the Intel machine, the dense-matrix
multiplication (DMM) and raytracer benchmarks have abundant, independent
parallelism and our compiler and runtime exploit them, demonstrating 
nearly ideal speedup over the baseline single-processor performance up to the
maximum number of cores.
Quicksort, barnes-hut, and spare-matrix multiplication (SMVM) all see reducing
speedups past 16~threads, but continue to steadily improve performance as more
threads are added.

On the AMD machine, shown in \figref{fig:amd-speedups}, DMM and the raytracer
benchmarks perform well.
But, both quicksort and barnes-hut scale nicely to 36~threads but then only take
slight advantage of additional threads. 
In barnes-hut, we believe that this behavior is due to the sequential portion.
Quicksort also is limited by its fork-join parallelism, and without
significantly increasing the size of the underlying dataset, it is difficult to
take advantage of the additional available parallelism.

Sparse-matrix multiplication provides the least scalability for the AMD system.
We believe that this is due to a large amount of available execution parallelism
but a relatively small amount of data.
Unless this data is either perfectly divided between the nodes or replicated to
each location, this benchmark fails to take much advantage of greater than even
24~threads.
We believe that the Intel machine's greater performance, particularly on SMVM,
is due to a smaller NUMA penalty when accessing the relatively smaller amount of
shared data, much of which resides on only one node.
Additionally, with only four nodes on the Intel machine, threads are twice as
likely to be located near data even if that data was placed randomly.

Benchmarks such as dense-matrix multiplication and raytracer, with excellent
locality and almost no shared data can scale nearly perfectly if all of their
data is kept locally.
The other benchmarks, which feature either heavily shared data or significant
points that sequentially merge data before creating more parallel work show
diminished improvements.
In all cases, poor locality negatively affects performance, particularly on
machines with multiple processor packages and relatively large numbers of
cores --- in our experience, between 24~and~36.

\subsection{Effect of allocation location}
By default, we allocate memory pages on the same node as the pinned \vproc{}
that required additional memory.
As a further test of locality, we modified the allocator for our garbage
collector with two alternative strategies that are similar to those of other
functional language single-threaded and parallel garbage collectors.
In \figref{fig:amd-speedups-interleaved}, we use an allocation strategy that
balances physical page assignments between the hardware packages.
This strategy is currently used in the Glasgow Haskell Compiler (GHC).
In \figref{fig:amd-speedups-single}, the allocation strategy defaults to a single
node for all allocations, which is the default NUMA behavior encountered by
single-threaded garbage collectors.
These speedup graphs are both plotted relative to the single-processor
performance for the AMD machine in \figref{fig:amd-speedups}.

Our strategy, which allocates pages local to the pinned \vproc{} that requests
and used the data, provides slightly better absolute performance at all
processor counts on all benchmarks except for SMVM in the interleaved strategy
at greater than 24~cores.
In that benchmark, there is a small portion of data (the vector) that is
accessed by all of the threads.
Our default implementation encounters bus saturation on the AMD machine at
larger numbers of processors, as all nodes are attempting to access data located
in the same package.

The single-node allocation strategy shows reasonable scalability until 12~cores.
But, this strategy fails after that point, and we expect all
collectors using this approach to require NUMA allocation tuning.\footnote{The
  current garbage collector for the Glasgow Haskell Compiler (GHC) recently
  required exactly this change in order to scale to even 7~processors across two
  sockets.}

\CUT{
\subsection{Load-balanced global collections}
\label{sec:evalGlobal}
During our global copying heap collection, there is often imbalanced available
work.
This imbalance results from uneven allocation patterns by threads,\footnote{In
  particular, when a large data file is sequentially read into memory.} and
causes there to be extra blocks of to-space available to scan.
Prior work on the Glasgow Haskell Compiler (GHC) showed increased times when
parallel collections also performed global load-balancing of this imbalanced
work~\cite{multicore-haskell}.
As we show in \tblref{eval:load-balanced}, load-balancing is very effective when
performed on a per-node basis.
We compare the performance of a unbalanced collections with balanced
collections on our AMD machine, where in the latter case threads will scan
unscanned chunks generated from any thread, so long as it is on the same node.
Execution times are measured in seconds, and lower numbers are better.
The DMM, SMVM, and raytracer benchmarks are not included because none of those
programs keep data around for long enough to trigger a global GC collection.
Barnes-hut and quicksort both see greater than a 15\% reduction in global GC
time and a nearly 3\% reduction in overall execution time, when run on
48~cores.
\begin{table}
  \begin{center}
  \begin{tabular}{r | c | c | c | c}
 & \multicolumn{2}{c|}{Unbalanced} & \multicolumn{2}{c}{Balanced} \\
Benchmark & Global (s) & Total (s) & Global (s) & Total (s)\\
\hline
Barnes-hut & 0.308 & 2.52 & 0.255 & 2.45 \\
Quicksort & 0.321 & 2.16 & 0.268 & 2.10 \\
% Barnes-hut & 82.8\% & 97.2\% \\
% Quicksort & 83.5\% & 97.2\% \\
\end{tabular}
\end{center}
\caption{
  Comparison of load-balanced versus unbalanced global collections on 48
  cores on the AMD machine. Smaller numbers are better.
}
\label{eval:load-balanced}
\end{table}%
}

\CUT{
\subsection{Single-thread performance versus a sequential program}

As is shown in \tblref{eval:mlton}, only on the raytracer do we offer
competitive performance to the sequential MLton baseline.
On all other benchmarks, we generally reach speed parity with four threads on
the AMD machine.
This performance gap is due to missed opportunities for sequential optimization
in Manticore and some small overhead from our parallel language constructs.
But, given that MLton has state-of-the-art functional language performance, this
comparison demonstrates that we have performance comparable with mainstream
functional languages on these benchmarks.

\begin{table}
  \begin{center}
  \begin{tabular}{r | c | c | c | c}
Benchmark & MLton (s) & 1T (s) & 2T (s) & 4T (s)\\
\hline
Barnes-hut & 20.4 & 61.4 & 38.2 & 17.0 \\
DMM & 12.4 & 50.6 & 27.6 & 12.6 \\
Quicksort & 20.9 & 100.8 & 52.5 & 27.5 \\
Raytracer & 10.9 & 15.6 & 7.7 & 3.9 \\
SMVM & 7.2  & 28.3 & 15.0 & 7.6\\
\end{tabular}
\end{center}
\caption{
  Comparison of Manticore performance versus MLton at low numbers of
  threads on the AMD machine. Smaller execution times are better.
}
\label{eval:mlton}
\end{table}%
}
%%% Local Variables: 
%%% mode: latex
%%% TeX-master: "paper"
%%% End: 

%\input{related}
%!TEX root = paper.tex
%
\section{Conclusion}
\label{sec:concl}

We have demonstrated a garbage collector designed to make effective use of the
memory hierarchy and that scales very well on a large number of processor cores. 
Keys to this design are private minor heaps that are collected concurrently with
program execution and in parallel with one another and a major heap architecture
that allows parallel collections while avoiding increasing traffic on the memory
bus.
Though some aspects of our system would need to be enhanced, for example with
write barriers or static analysis, in the context of systems that permit and
encourage frequent unrestricted memory mutation, we believe that these techniques
are readily applicable to other runtimes. 

\paragraph{Acknowledgments}
Thanks to Bradford Beckmann for reviewing the breakdown of the AMD G34 socket.
This material is based upon work supported by the National Science Foundation
under Grants CCF-0811389 and CCF-1010568.
The views and conclusions contained herein are those of the authors and should
not be interpreted as necessarily representing the official policies or
endorsements, either expressed or implied, of these organizations or the
U.S.\ Government.

Access to the Intel machine was provided by Intel Research.
Thanks to the management, staff, and facilities of the Intel Manycore Testing
Lab.\footnote{Manycore Testing Lab Home:\\
  \url{http://www.intel.com/software/manycoretestinglab}\\
Intel Software Network:\\
\url{http://www.intel.com/software}}

\bibliographystyle{common/alpha}
\bibliography{common/strings-short,common/manticore}

\newpage
\appendix
%!TEX root = paper.tex

\section{Hardware}
\label{sec:hardware}

% Your system supports ECC DDR3 registered DIMMs (RDIMMs) and unbuffered
% registered DIMMs (UDIMMs). Quad-rank DIMMs of 1066 MHz and single- and
% dual-rank DIMMs of 1333 MHz are also supported. RDIMMs of capacities 2
% GB, 4 GB, and 8 GB are supported for a total of up to 256 GB.

% The system consists of 32 memory sockets split into four sets of eight
% sockets; one set for each processor. Each eight-memory socket set is
% further organized into four DDR3 memory channels. The first socket of
% each DDR3 memory channel is marked with a white release lever.

% Each channel supports up to two single-, dual-, or quad-rank RDIMMs or
% two UDIMMs. The interface uses either 2 GB, 4 GB, or 8 GB RDIMMs and 1
% GB, 2 GB, or 8 GB UDIMMs.

% The following features are available with respect to memory:
% * Each processor has four DDR3 channels that support speeds up to 1333 MHz.
% * UDIMMs and RDIMMs cannot be mixed.
% * Quad-Rank DIMM types support speeds up to 1066 MHz.
% * Your system may support online memory sparing. 
% * Mixing of memory sizes and ranks is allowed for flexibility.
% * Up to 256 GB of memory (with 32 8 GB RDIMMs)
\subsection{AMD Hardware}
\label{intro:amdhardware}
\begin{figure}
\centering
\includegraphics[scale=0.5]{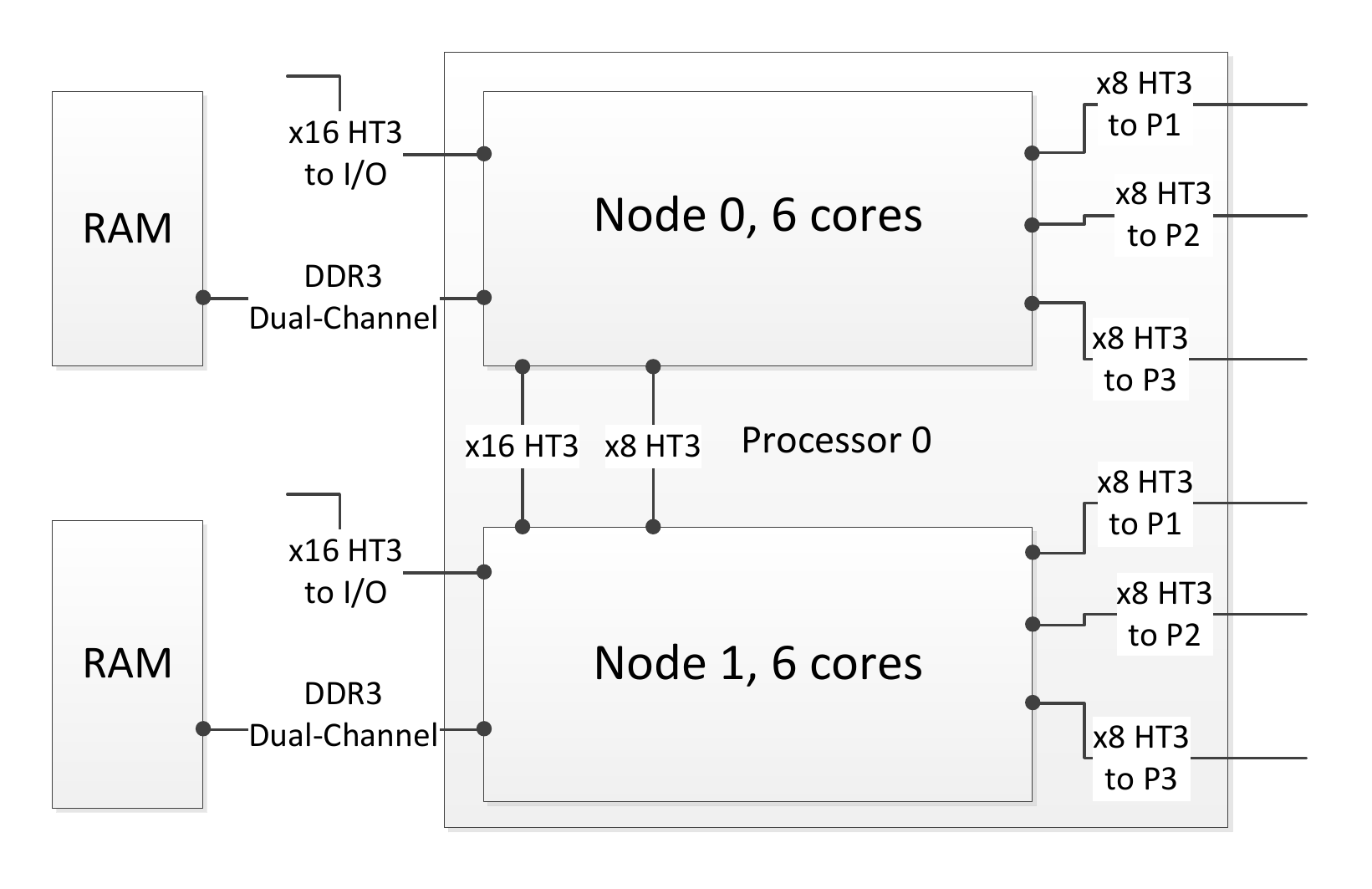}
\caption{
Interconnects for one processor in a quad AMD Opteron machine.
}
\label{amd:magny}
\end{figure}

Our AMD benchmark machine is a Dell PowerEdge R815 server, outfitted
with 48~cores and 128~GB physical memory.
This machine runs x86\_64 Ubuntu Linux 10.04.2 LTS, kernel version 2.6.32-27.
The 48~cores are provided by four AMD Opteron 6172 ``Magny Cours'' processors~\cite{magny-cours,opteron},
each of which fits into a single G34 socket.
Each processor contains two nodes, and each node has six cores.
The 128~GB physical memory is provided by thirty-two 4~GB dual ranked RDIMMs,
evenly distributed among four sets of eight sockets, with one set for each processor.
As shown in \figref{amd:magny}, these nodes, processors, and RAM chips form a
hierarchy with significant differences in available memory bandwidth and number
of hops required, depending upon the source processor core and the target physical memory location.
Each 6~core node (die) has a dual-channel double data rate 3 (DDR3) memory
configuration running at 1333~MHz from its private memory controller to its own
memory bank.
There are two of these nodes in each processor package. 

Bandwidth between each of the nodes and I/O devices is provided by four 16-bit
HyperTransport 3 (HT3) ports, which can each be separated into two 8-bit HT3
links.
Each 8-bit HT3 link has 6.4~GB/s of bandwidth. 
The two nodes within a package are configured with a full 16-bit link and an
extra 8-bit link connecting them. 
Three 8-bit links connect each node to the other three packages in this four
package configuration. 
The remaining 16-bit link is used for I/O.
\tblref{numa:bandwidth} shows the bandwidth available between the different
elements in the hierarchy.
\begin{table}
  \begin{center}
  \begin{tabular}{r | c | c}
  \multicolumn{1}{c|}{} & AMD (GB/s) & Intel (GB/s)\\
\hline
Local Memory & 21.3 & 17.1\\
Node in same package & 19.2 & \emph{n/a}\\
Node on another package & 6.4 & 25.6\\
\end{tabular}
\end{center}
\caption{
  Theoretical bandwidth available between a single node and the rest of the system.
}
\label{numa:bandwidth}
\end{table}%

Each core operates at 2.1~GHz and has 64~KB each of instruction and data L1 cache and 512~KB of L2 cache.
Each node has 6~MB of L3 cache physically present, but, by default, 1~MB is
reserved to speed up cross-node cache probes.

%% Short answer the MTL consists of 4 socket Intel(R) Xeon X7560 (with
%% 8-cores/16-threads)

%% See: http://ark.intel.com/Product.aspx?id=46499 for more details.

%% The MTL is a white-box enterprise server, but the current OEM (QSSC) provides a 
%% detailed Technical Product Spec (that matches the white-box configuration) on
%% this page: 
%% http://www.qsscit.com/en/01_product/03_download.php?page=10&mid=27&sid=125&id=126&qs=50
% http://www.qsscit.com/language_config/down.php?hDFile=S4R_TPS_1.0.pdf

%% The memory (riser) boards are populated with 4GB DDR3 1066 MHz registered
%% DIMMs.
% Total: 264147540

% 24MB L3 cache!
\subsection{Intel Hardware}
\label{intro:intelhardware}
\begin{figure}
\centering
\includegraphics[scale=0.5]{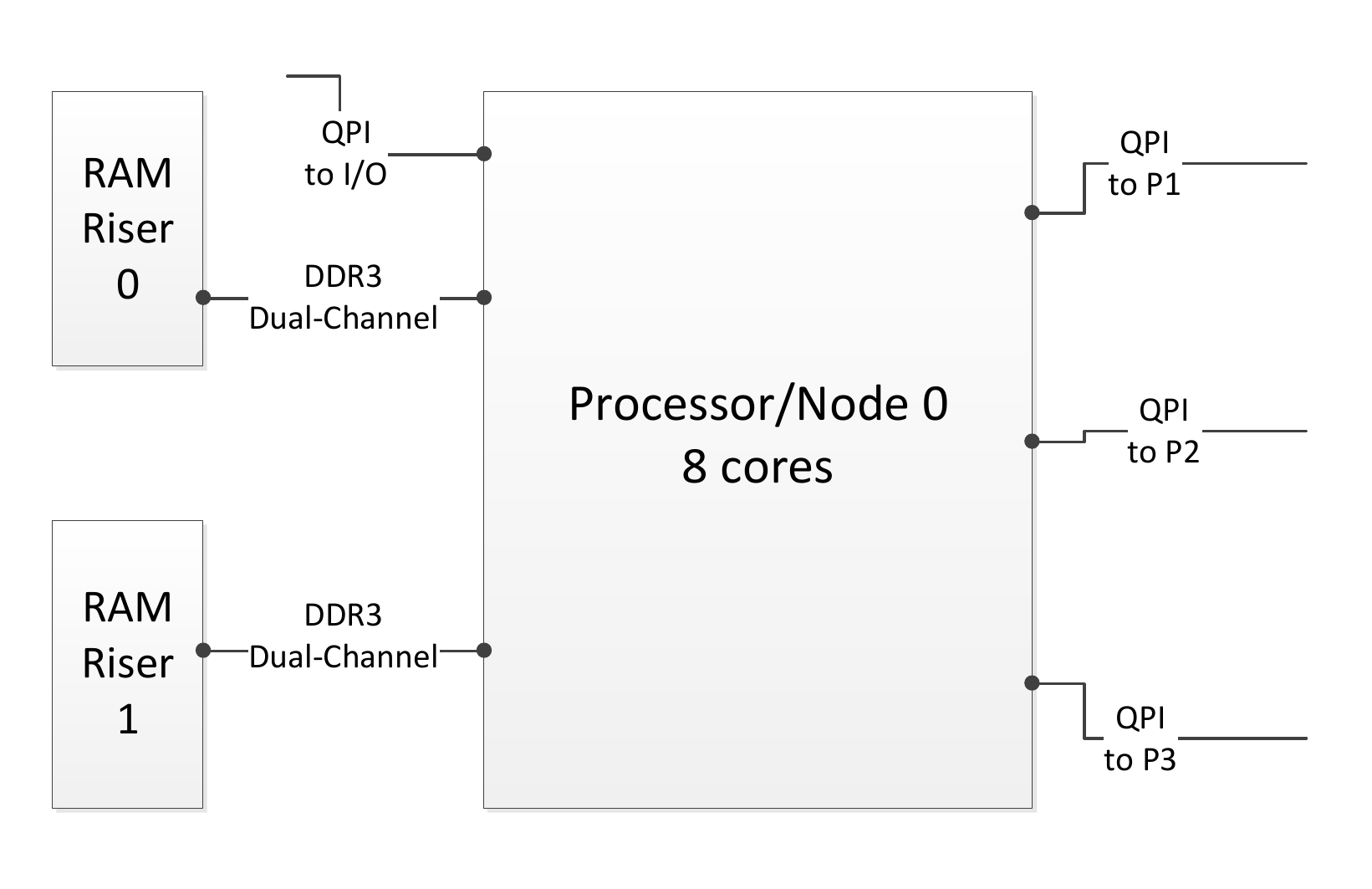}
\caption{
Interconnects for one processor in a quad Intel Xeon machine.
}
\label{intel:xeon}
\end{figure}

The Intel benchmark machine is a QSSC-S4R server with 32~cores and 256~GB
physical memory.
This machine runs x86\_64 RedHat Enterprise Linux, kernel version
2.6.18-194.11.4.el5.
The 32~cores are provided by four Intel Xeon X7560 processors~\cite{xeon,qssc}.
Each processor contains 8~cores, which can be but are not configured to run with
2~simultaneous multithreads (SMT).
As shown in \figref{intel:xeon}, these nodes, processors, and RAM chips form a
hierarchy, but this hierarchy is more uniform than that of the AMD machine.

Each of the nodes is connected to two memory risers, each of which has a
dual-channel DDR3 1066~MHz connection.
The 4~nodes are fully connected by full-width Intel QuickPath Interconnect (QPI)
links.
\tblref{numa:bandwidth} shows the bandwidth available between the different
elements in the hierarchy.

Each core operates at 2.266~GHz and 32~KB each of instruction and data L1 cache
and 256~KB of L2 cache.
Each node has 24~MB of L3 cache physically present but, by default, 3~MB is
reserved to speed up both cross-node and cross-core caching.

%%% Local Variables: 
%%% mode: latex
%%% TeX-master: "paper"
%%% End: 

\end{document}